\documentclass[lettersize,journal]{IEEEtran}
\usepackage{amsmath,amsfonts}
\usepackage{algorithmic}
\usepackage{algorithm}
\usepackage{array}
\usepackage[caption=false,font=normalsize,labelfont=sf,textfont=sf]{subfig}
\usepackage{textcomp}
\usepackage{stfloats}
\usepackage{url}
\usepackage{verbatim}
\usepackage{graphicx}
\usepackage{cite}
\usepackage{booktabs}
\usepackage{hyperref}
\usepackage{mathptmx}                  

\usepackage{amssymb}
\usepackage{enumitem}
\usepackage[normalem]{ulem}

\usepackage{tikz}
\usepackage{xcolor}
\usepackage[scale=0.82]{FiraMono}

\definecolor{maroon}{RGB}{138, 16, 11}
\definecolor{gray}{RGB}{114, 97, 88}
\definecolor{gold}{RGB}{220, 202, 160}

\begin{document}

\title{Understanding the Research-Practice Gap in Visualization Design Guidelines}

\author{Nam Wook Kim, Grace Myers, Jinhan Choi, Yoonsuh Cho, Changhoon Oh, Yea-Seul Kim
\thanks{N. W. Kim, G. Myers, J. Choi, and Y. Cho are with the Department of Computer Science, Boston College, Chestnut Hill, MA 02467 USA (e-mail: \{nam.wook.kim, grace.myers, jinhan.choi, yoonsuh.cho\}@bc.edu).}%
\thanks{C. Oh is with the Graduate School of Information, Yonsei University, Seoul, South Korea (e-mail: changhoonoh@yonsei.ac.kr).}%
\thanks{Y. Kim is with the Department of Computer Sciences, University of Wisconsin--Madison, Madison, WI 53706 USA (e-mail: yeaseul.kim@cs.wisc.edu).}%
}

\markboth{Journal of \LaTeX\ Class Files,~Vol.~14, No.~8, August~2021}%
{Shell \MakeLowercase{\textit{et al.}}: A Sample Article Using IEEEtran.cls for IEEE Journals}


\maketitle

\begin{abstract}
Although empirical research often underpins practical visualization guidelines, it remains unclear how well these research-driven insights are reflected in the guidelines practitioners actually use. In this paper, we investigate the research-practice gap in visualization design guidelines through a mixed-methods approach. We collected 390 design guidelines from practitioner-facing sources and 235 empirical studies to quantitatively assess their alignment. To complement this analysis, we conducted surveys with 69 participants (33 practitioners, 36 researchers) and in-depth interviews with 20 experts to examine their experiences, perceptions, and challenges. Our findings reveal discrepancies: empirical evidence often contradicts or only partially supports widely used guidelines, and the two communities prioritize different attributes of design. Based on these insights, we derive a holistic guideline template (integrating \textit{Context}, \textit{Approach}, \textit{Problem}, and \textit{Purpose}) and discuss actionable strategies, such as a triadic knowledge model.

\end{abstract}

\section{Introduction}

Empirical research has provided foundational knowledge for data visualization. Graphical perception studies, beginning with the seminal work by Cleveland and McGill~\cite{cleveland1984graphical}, have provided useful insights into which visual channels are more perceptually effective. Visual cognition studies have contributed to a better understanding of the higher-level processing of visualizations, including comprehension~\cite{bateman2010useful}, engagement~\cite{borgo2012empirical}, memorability~\cite{borkin2015beyond}, and biases~\cite{dimara2018task}. Many of these empirical findings have been distilled into design guidelines that are now widely referenced in practitioner resources~\cite{choi2023vislab}; examples include using bar charts instead of pie charts for comparing values or including a zero baseline. While empirical research continues to expand and refine such guidelines, it remains unclear how well research-driven guidance is reflected in the guidelines practitioners actually use.

In this study, we investigate the research-practice gap in visualization design guidelines, aiming to understand their interplay better and promote evidence-based design practices. As the field of visualization increasingly influences decision-making across society, it is crucial to bridge this gap to help practitioners make informed choices for creating truthful and functional visualizations. Our approach mirrors initiatives in fields such as medicine, education, and policy-making, where the shift has been from relying on tradition, intuition, and anecdotal experience to making decisions based on solid research findings~\cite{evidpractice,bogenschneider2011evidence,davies1999evidence,sackett1996evidence}. This movement also relates to intermediary efforts, often referred to as translational science~\cite{woolf2008meaning}, which focus on converting research results in research into practical implementations~\cite{norman2010research}. Although translation can go in both directions, our focus is specifically on examining the design guidelines accessible to practitioners and assessing how well they align with empirical research findings.

Our specific research questions include:
\begin{itemize}
    \item What design guidelines are currently accessible to practitioners and how do these align with empirical visualization research?
    \item How do practitioners and researchers perceive the research-practice gap in visualization design guidelines? 
\end{itemize}

We employed a two-phase mixed-methods approach to explore these questions. We first collected data visualization \textit{design guidelines} in the wild by conducting a keyword-based web search, and then conducted a qualitative content analysis to characterize their attributes. Our findings revealed a concentration of guidelines on basic charts and task types and identified three distinctive categories of the guidelines: \textit{problem-solving}, \textit{enhancement}, and \textit{alternative suggestion}. To understand the agreement and discrepancies between guidelines and empirical research, we analyzed how they are aligned with each other; we focused on empirical studies that evaluate user performance across various visualization designs~\cite{lam2011empirical}. The alignment analysis revealed inconsistencies between mapped guidelines and their associated empirical evidence, highlighting a disconnect between research findings and practitioner-facing guidelines. Unmapped studies suggest areas for new guidelines, while unsupported guidelines offer opportunities for empirical research.

The content analysis of the guideline landscape provided insights into the available guideline knowledge in practice, but it did not capture the lived experiences and perspectives of researchers and practitioners regarding the knowledge gaps. To gain a more comprehensive understanding, we supplemented this with surveys and interviews. Practitioners were mainly recruited from the Data Visualization Society~\cite{datavissociety}, while researchers were recruited through community mailing lists. A total of 33 practitioners and 36 researchers filled out our survey questionnaire on guidelines, empirical studies, and the challenges of transferring knowledge between the two communities. The survey results revealed similarities and differences in their access to guidelines, perceptions of their usefulness, and the challenges they encountered, while also indicating a shared recognition of the gap in transferring research into design guidelines~\cite{parsons2021understanding}. In follow-up interviews with ten selected participants from each community, we gained deeper insights into their experiences and struggles with guidelines, as well as their attitudes and perceived causes of the knowledge gap and ideas for addressing it. 

Our primary contribution lies in enhancing the understanding of the design guideline knowledge gap between research and practice, paving the way for evidence-informed design decision-making. We synthesized our findings into a range of strategies and considerations, detailed in the discussion section, aimed at bridging the guideline-focused knowledge divide. These strategies include establishing consistent vocabularies for data visualizations, creating a triadic model that integrates guidelines, empirical evidence, and contextual examples, and improving guideline accessibility through detailed metadata annotation and template development. We also highlight the importance of fostering collective efforts in knowledge brokerage~\cite{ward2009knowledge}, essential for the growth and evolution of the field.

\section{Background: Research, Practice, \& Education}
The field of data visualization has always been closely tied to practical needs and applications. Widely used software tools and programming toolkits, such as Tableau~\cite{tableau,stolte2002polaris}, D3~\cite{d3}, and Vega-Lite~\cite{vegalite}, began as research projects, as did visualization techniques such as treemaps~\cite{johnson1998tree} and parallel coordinates~\cite{inselberg1990parallel}. Researchers have also devoted much effort to understanding the challenges and struggles faced by practitioners in deriving new research insights~\cite{joyner2022visualization,bigelow2014reflections} and building new visualization tools for practitioners~\cite{kim2016data,liu2018data}. Furthermore, the growing sub-fields of visual analytics~\cite{cook2005illuminating} and design studies~\cite{sedlmair2012design} directly address problems faced by domain experts across disciplines.

Similarly, the visualization research community has developed theories and frameworks over the past few decades that provide established vocabularies for reading and creating visualizations, such as visual encoding channels, data types, and analytical tasks~\cite{brehmer2013multi,munzner2014visualization}. To evaluate the usefulness of a visualization design, researchers conduct experiments that compare design alternatives. The results of these experiments provide guidance for selecting effective visual encoding channels~\cite{cleveland1984graphical, heer2010crowdsourcing} and determining their effectiveness for different tasks~\cite{kim2018assessing,saket2018task}. Some of these results are integrated into existing tools that automate visual encoding design~\cite{tableau, voyager, hu2019vizml}. Although these automated tools suffice for exploratory analysis, designers must still consider a range of design factors to ensure the successful and engaging communication of data, such as selecting appropriate color schemes and determining axis ranges.

Researchers have developed pedagogical approaches to teaching data visualizations, such as the Five Design-Sheet Methodology~\cite{roberts2015sketching} and VisItCards~\cite{he2016v}. Other studies have examined data visualization literacy, which is defined as the ability to read information encoded in data visualizations and create them to answer questions about the data~\cite{lee2016vlat,borner2019data,alper2017visualization,boy2014principled,ge2023calvi}. 
However, these works in pedagogy and literacy tend to prioritize higher-level design processes and chart-reading tasks, often placing less emphasis on practical knowledge like actionable design guidelines.

Several recent studies have attempted to address more practical design needs. Diehl et al. launched VisGuides, a forum that provides practitioners with a central place to ask questions about the validity and quality of their visualizations and discuss guidelines around visualization design~\cite{diehl2018visguides}. Wang et al. proposed a cheatsheet structure to introduce a visualization technique, including its construction process, similar techniques, and potential issues~\cite{wang2020cheatsheet}. On the other hand, researchers have discussed theoretical models for connecting with empirical studies~\cite{brown2018tagguideline} with guidelines and using guidelines to describe relationships between various stages of the design process~\cite{meyer2014nestedblock}. 

Similar efforts have also been made by practitioners as well. Resources such as \textit{From Data to Viz} and \textit{DataVizProject} provide user interfaces to select charts based on different data types and tasks~\cite{datatoviz,datavizproject}. Companies publish design systems that often include recommendations on visualization design~\cite{adobedesignsystem,googledesignsystem}, although they mostly focus on styling guidelines to match their brands.

Currently, visualization design guidelines are dispersed across different resources, often through personal blogs and published books. It remains unclear how, or even if, practitioners learn critical design knowledge from these guidelines. Furthermore, researchers often convey research outcomes through blogs~\cite{multipleviews}, podcasts~\cite{datastories}, or social media in an ad-hoc manner. While recent studies have begun investigating the gap between research and practice, they mostly focus on understanding holistic design processes rather than acquiring fundamental data visualization design knowledge~\cite{parsons2021understanding,zhang2022visualization}.

This paper aims to understand the design guideline knowledge gap between visualization design practice and research. Bridging this gap is essential because it enables practitioners to leverage the latest research to make more informed decisions, resulting in better data visualizations. In addition to fields including medicine and biology, people in a related field of human-computer interaction, have recognized this issue as well and studied how to better apply research findings to practice, a process known as translational research~\cite{woolf2008meaning}.

For instance, researchers conducted interviews with practitioners to understand their barriers to research findings~\cite{colusso2017translational} and to propose new models to tackle the research-practice gap~\cite{colusso2019translational,beck2018theory}, mostly focusing on the domain of user experience design. Other researchers have examined how specific research frameworks are applied by user experience designers~\cite{velt2020translations,remy2015bridging}. Other studies have focused on specific subdomains such as AI in UX~\cite{yildirim2023investigating}, human-robot interaction~\cite{cila2024bridging} interaction design for children~\cite{kawas2021translational}. 

We focus on data visualization which is uniquely different from these fields. The design of effective visualizations can be grounded in perceptual and cognitive studies through systematic investigations based on a well-defined design space, including encoding, data types, and tasks, offering a distinct perspective on alleviating the gap between research and practice. In this work, we employ a range of mixed methods, including content analysis, surveys, and interviews, engaging both practitioners and researchers, to investigate the knowledge gap in guidelines.



\section{Guideline and Empirical Study Alignment}   
To better understand the research-practice gap in visualization design guidelines, we first examined how existing practitioner-facing guidelines align with empirical research findings. This analysis helps reveal whether current guidelines are well-supported by research evidence, identify areas of mismatch, and uncover opportunities for new guideline development. 


\subsection{Data Collection}
\subsubsection{Collecting guidelines}
\label{sec:collecting-guidelines}
To ensure comprehensiveness in our collection, we extensively consulted a variety of guideline resources in practical settings, such as industry resources and online platforms. We first used web search by using a combination of different keywords, such as \textit{data visualization}, \textit{guidelines}, and similar synonyms, and went through up to five pages of the top results. We also looked at known resources such as blogs (e.g., DataWrapper~\cite{datawrapper}) and websites (e.g., from Data to Viz~\cite{datatoviz}, Data Journalism~\cite{datajournalism}). 
Finally, we consulted the guideline collection curated by the Data Visualization Society~\cite{datavissociety}, which provided additional blogs, websites, and books (e.g., How Charts Lie by Cairo~\cite{cairo2019charts}). Our search concluded once we reached a saturation of overlapping and duplicate guidelines, totaling 1,279 guidelines.

Next, we applied inclusion and exclusion criteria to serve our purpose of analyzing visualization guidelines in practice. For instance, we only considered guidelines relevant to data visualization design, adopting the definition from Chen et al.~\cite{chen2017pathways}; i.e., a guideline describes \textit{a process or a set of actions that may lead to a desired outcome or, alternatively, actions to be avoided to prevent an undesired outcome.} Consequently, we excluded guidelines offering narrow styling advice, such as increasing font size. We also disregarded broader recommendations, such as making visualizations interactive. Similarly, we excluded generic chart suggestions, such as using a bar chart for a single series data, that are too evident to be considered guidelines. We also consolidated guidelines that significantly overlapped in content or purpose. After the filtering and aggregation process, we obtained the final collection of \textbf{390 guidelines} from 32 diverse sources.  Of these, 49\% were derived from blog articles, 39\% from style guidelines, and the remaining 12\% from books.

\paragraph{Why we did not consider guidelines from research papers}

We focused on practitioner-facing sources to capture the guidelines actually consumed by the target audience, distinguishing them from the academic source material. Findings from Section \ref{sec:surveys-interviews}), along with prior literature~\cite{colusso2017translational}, confirm that practitioners rarely consult academic research papers.

\paragraph{Why we did aggregated similar guidelines}
We aimed to avoid overestimating the results of our alignment analysis, which examines the extent to which guidelines align or mis-align with empirical research by giving equal weight to each guideline. However, analyzing the overlap in the frequency of each guideline could reveal which ones are most commonly used in practice, though this is not the primary focus of this paper.

\paragraph{Why we only considered certain types of guidelines}
Design knowledge is a broad concept that encompasses anything related to design. To narrow our scope, we focused on guidelines using the definition provided by Chen et al. However, guidelines themselves can also be broadly defined, so we refined our focus to those that contain meaningful design knowledge that is specific to data visualization. For instance, a chart type predetermines the kinds of data that can be represented, and thus it may not require empirical support. On the other hand, while guidelines can be based on experience, they may still warrant empirically-oriented explanations (e.g., making a title engaging ~\cite{borkin2015beyond}).


\subsubsection{Collecting empirical studies}

We used a combination of automated and manual searches to collect empirical papers. First, we initially used a keyword search (e.g., visualization, evaluation, study) on Scopus to collect papers from ACM CHI and EuroVis over the past decade. We collected the same data for IEEE VIS from VisPub~\cite{isenberg2016vispubdata}. Expanding on this initial collection, we supplemented it with recent papers published in ACM CHI, IEEE VIS, and Eurovis from 2021 to 2022 to ensure the dataset reflects the latest research. We also manually included relevant past papers that were not captured by our initial search from additional sources, including \href{http://visperception.com/}{VisPerception}~\cite{visperception} and a recent survey of perception-based visualization studies~\cite{quadri2021survey}; this process aided in gathering influential works published prior to 2011, such as the notable graphical perception experiment conducted by Cleveland \& McGill~\cite{cleveland1984graphical}. Finally, two researchers cleaned up the data collection by removing any papers that did not present original empirical research and by only including studies that evaluate user performance on different visualization designs based on the characterization in a prior work~\cite{lam2011empirical} i.e., we excluded algorithm performance and system usability studies. The final collection of empirical studies consists of \textbf{235 research papers}, spanning from 1911 to 2022. A detailed breakdown of the papers by venues and years is available in the supplementary material.

\subsection{Analyzing Process}
\paragraph{Design guideline content analysis} To facilitate content analysis~\cite{potter1999rethinking} of the guidelines, two junior researchers compiled short descriptions from guideline articles into a spreadsheet. A lead researcher initially reviewed these descriptions and, when necessary, consulted the source articles. This process was to establish initial codes and code values, including chart types, data types, task types, and chart elements. 
The two junior researchers then independently coded the data using the initial codebook. They expanded it as necessary, adding new codes related to purpose, problem, approach, and specific features like color, animation, 3D, and interaction, consulting with the lead researcher to resolve any questions or conflicts in coding. Lastly, the lead researcher thoroughly reviewed the final codes. 

\paragraph{Empirical study content analysis} We derived the one-sentence purpose of each study out of its abstract using GPT4 to facilitate the following alignment process. We initially attempted to codify aspects such as chart types and experiment methods. However, given the comprehensive empirical study landscape already explored in recent studies~\cite{kosara2016empire,zeng2023review,quadri2021survey}, we ultimately decided not to perform a detailed codification of the empirical studies. This approach allowed us to maintain a high-level analysis that aligns with our study's strategic objectives, ensuring a focused and manageable research endeavor. Nonetheless, exploring finer-grained, code-level alignment in future work beyond our current entity-level alignment would be beneficial.

\paragraph{Aligning two corpora}
The alignment process followed a similar procedure to the content analysis. 
The lead researcher established exemplary mappings, and the remaining mapping tasks were distributed to five researchers (three are co-authors in this paper). We created a new spreadsheet for aligning guidelines with empirical studies, incorporating key information from each. This information included short descriptions, purpose, problem, and approach for guidelines, and the title, purpose, and abstract for empirical studies. We then mapped each guideline to relevant empirical studies, noting whether they were supported, contradicted, or yielded mixed evidence. Consultations with the original guideline sources and research articles were conducted as needed. Through this process, we observed that a guideline could correspond to multiple research papers, and vice versa.



\begin{figure*}[tb]
    
    \centering
    \includegraphics[width=\linewidth]{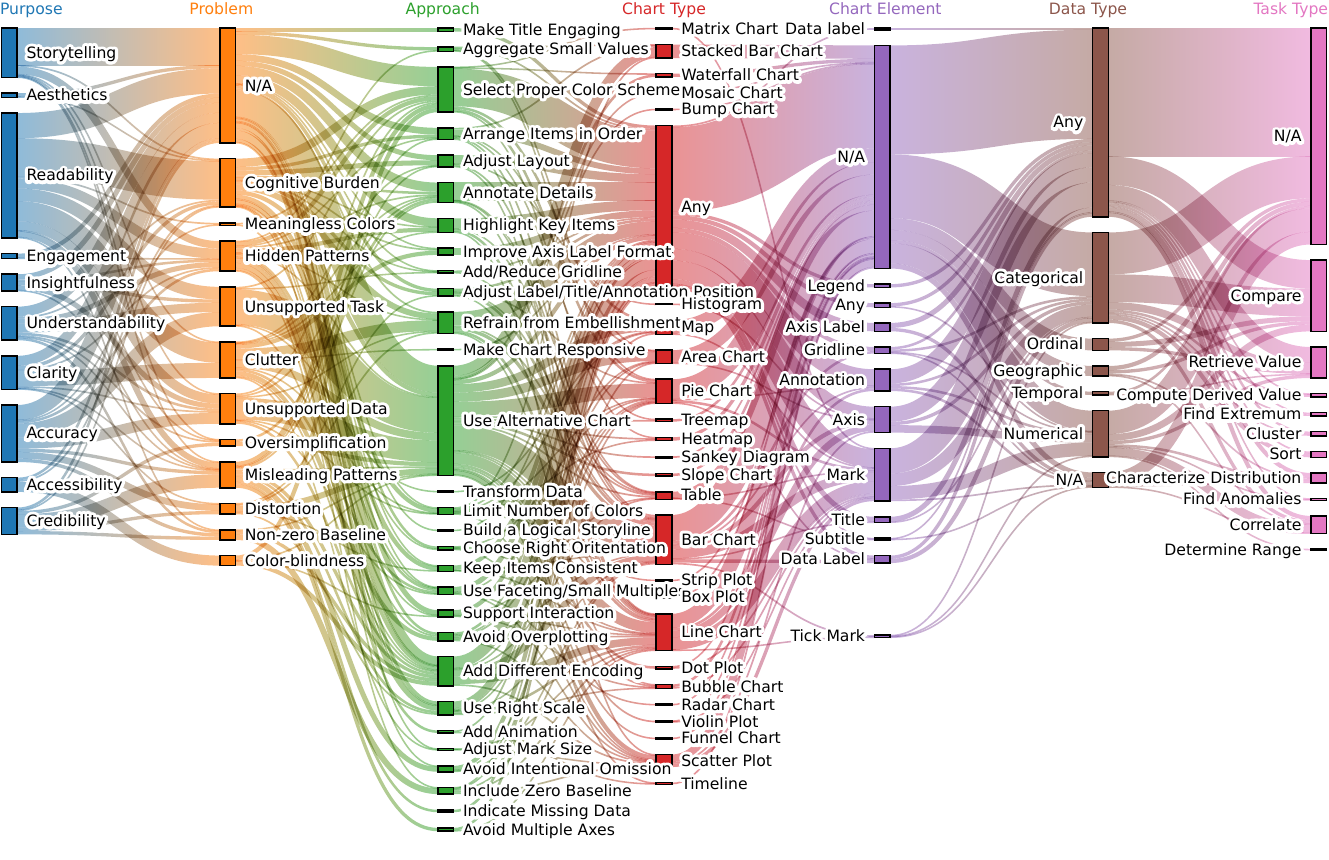}
    
    \caption{A Sankey diagram showing our final collection of 390 guidelines, revealing the proportional relationship among six attributes of visualization design guidelines---context (\textit{purpose}, \textit{problem}, \textit{approach}. \textit{chart type}, \textit{chart element}, \textit{data type}.}
    \label{fig:guideline-landscape}
    
\end{figure*}

\subsection{Guideline Landscape}

We identified primary high-level codes such as \textit{purpose}, \textit{problem}, \textit{approach}, as well as additional low-level codes, including \textit{chart type}, \textit{chart elements}, \textit{data type}, and \textit{task type}. \autoref{fig:guideline-landscape} shows an overview of our coding results.

\paragraph{Purpose}
Each guideline serves a specific purpose, representing a goal it aims to achieve. Our analysis identified various key objectives for these guidelines, including \textit{readability} (35\%), \textit{storytelling} (15\%), \textit{accuracy} (14\%), and \textit{clarity} (9\%). Readability refers to how easy it is to read text or visual elements; e.g., \textit{when working with large or dense data sets, avoid dot plots}. Storytelling purposes include adding additional context of the chart to improve explanation, such as \textit{making titles concise and engaging}. Accuracy pertains to a more precise perception of data; e.g., \textit{for a more accurate comparison, consider using a bar chart rather than a radial bar chart}. Clarity refers to how information is clearly expressed; e.g., \textit{When detail isn't needed, summarize}. 

Other purposes we observed include \textit{understandability} (8\%), \textit{credibility} (8\%), \textit{insightfulness} (5\%), \textit{accessibility} (4\%), \textit{engagement} (1\%), and \textit{aesthetics} (1\%).  Understandability concerns more about comprehension; e.g., \textit{when creating a scatterplot, show a trend line}.  On the other hand, credibility relates to how well the chart can be trusted; e.g., \textit{To prevent a false impression of the data, do not cherry-pick data.}. Other examples for insightfulness, accessibility, and engagement \& aesthetics, and include: \textit{when drawing a scatter plot, add marginal distributions to detect the distribution hidden in the overplotted parts of the graphic}; \textit{for colorblind people, color only the most important values}; \textit{to establish unity and cohesion, align all elements of the graph to create clean vertical and horizontal lines}.

We aimed to select the most representative purpose for each guideline, though these purposes often overlapped. For example, concepts like \textit{readability}, \textit{understandability}, and \textit{clarity} are similar and can be used interchangeably, yet they each focus on distinct aspects such as choosing, augmenting, or modifying a chart. Similarly, \textit{insightfulness} and \textit{credibility} may intersect with these metrics. Another group with a related spirit includes \textit{storytelling}, \textit{engagement}, and \textit{aesthetics}.

\paragraph{Problem}
We observed 12 common problems discussed in our guideline collection. Notable problems include \textit{cognitive burden} (15\%), \textit{clutter} (9\%), \textit{hidden patterns} (9\%), \textit{unsupported task} (8\%), \textit{misleading patterns} (8\%), and \textit{unsupported data} (7\%). For instance, \textit{cognitive burden}, involving issues like inconsistency, illegibility, and ambiguity, can lead to eyestrain and attention fatigue. An example would be reducing back and forth eye movement by positioning data labels near their corresponding data points. Moreover, these problems often reflect commonly recognized issues in the field. For instance, employing choropleth maps for detecting subtle regional differences can lead to an issue classified as an \textit{unsupported task}. Additionally, using transparency is advisable when dealing with hidden or overlapping data points, helping to reveal \textit{hidden patterns}.

Other problems focused on more specific issues, such as concerning \textit{distortion} (3\%), \textit{color-blindness} (3\%), \textit{non-zero baseline} (3\%), \textit{oversimplification} (2\%), and \textit{meaningless colors} (1\%). Example guidelines include avoiding inverted axes to prevent \textit{distortion}; incorporating shapes and textures alongside color to aid those who are \textit{colorblind}; avoid giving each bar a categorical color to deter \textit{meaningless colors.} These issues, while potentially fitting under broader categories like cognitive burden, stood out enough to be classified individually. Interestingly, not all guidelines were problem-focused (33\%), with some merely suggesting improvements (\autoref{fig:template}).

\paragraph{Approach}
We noticed the utilization of 29 distinct approaches aimed at tackling design issues or improving the situation, highlighting considerable diversity in the methods employed. The list of top three approaches includes: \textit{use alternative chart} (24\%), \textit{select proper color scheme} (14\%), and \textit{add different encoding} (8\%). Most of these approaches concentrate on selecting effective visual encodings. One strategy is to opt for alternative chart types, as exemplified by a guideline that advocates \textit{using a connected scatterplot instead of a dual-axis chart when merging two line charts}. Another method is the use of varied mark types and encoding channels. For instance, \textit{when red and green are used together, adding symbols as an auxiliary communication method is recommended}. 

Color was also a prominent issue, with 98 out of 390 guidelines pertaining to color. The category \textit{select proper color scheme} encompassed guidelines addressing rainbow color palettes, color blindness, highlighting, low-contrast issues, and inconsistent color semantics, such as using darker colors for lower values. The top ten list of approaches continues with \textit{refrain from embellishment} (6\%), \textit{annotate details} (6\%), \textit{highlight key items} (4\%), \textit{use right scale} (4\%), \textit{adjust layout} (4\%), \textit{arrange items in order} (4\%), and \textit{avoid overplotting} (3\%). \autoref{fig:guideline-landscape} shows the full list of remaining approaches.

\paragraph{Types of Charts, Data, Tasks}

A large portion of the guidelines (40\%) were not specific to particular chart types classified as \textit{any} in \autoref{fig:guideline-landscape}. When specificity was provided, it centered largely on standard charts like bar charts (11\%), line charts (9\%), pie charts (6\%), scatter plots (4\%), and choropleth maps (2\%).

Many of the other guidelines were related to variations of these standard chart types, such as stacked bar charts (4\%), area charts (4\%), and bubble charts (1\%). Oftentimes, guidelines were focused on specific chart elements (42\%). \textit{Mark} was the most frequently discussed topic (15\%), followed by \textit{axis} (8\%), \textit{annotation} (7\%), and \textit{axis labels} (3\%). As an example, the guidelines for \textit{mark} encompass suggestions such as \textit{keeping the distance between the bars at half the width of the bars to enhance data comprehension} and \textit{if a bar chart contains numerous bars causing clutter, consider using a dot plot as circles consume less ink than bars}. An illustrative guideline under \textit{axis} recommends that \textit{when creating time-series plots with unevenly collected data, ensure that the tick marks on the time axis reflect the irregular intervals to prevent distortion of the information}.

We endeavored to code data types using the well-established framework in the research community, which includes categorical, numerical, temporal, and geospatial data types. However, we observed that design guidelines rarely specify such data types (\autoref{fig:guideline-landscape}), often resulting in unreliable codes. Such data types could often be inferred from the chart types, however. We used Amar et al.\cite{amar2005low}'s taxonomy of low-level analytic tasks to code task types. A significant portion of the 150 guidelines (38\%), where task types were available, centered around the basic tasks of \textit{compare} (18\%) and \textit{retrieve value} (7\%); \textit{compare} is a combination of \textit{retrieve value} and \textit{compute derived value}. Other task types observed include \textit{correlate} (5\%), \textit{characterize distribution} (3\%), and \textit{sort} (2\%).

\subsection{Guideline Templates}

\label{sec:guideline-landscape}
\begin{figure*}[th]
    \centering
    \includegraphics[width=0.70\linewidth]{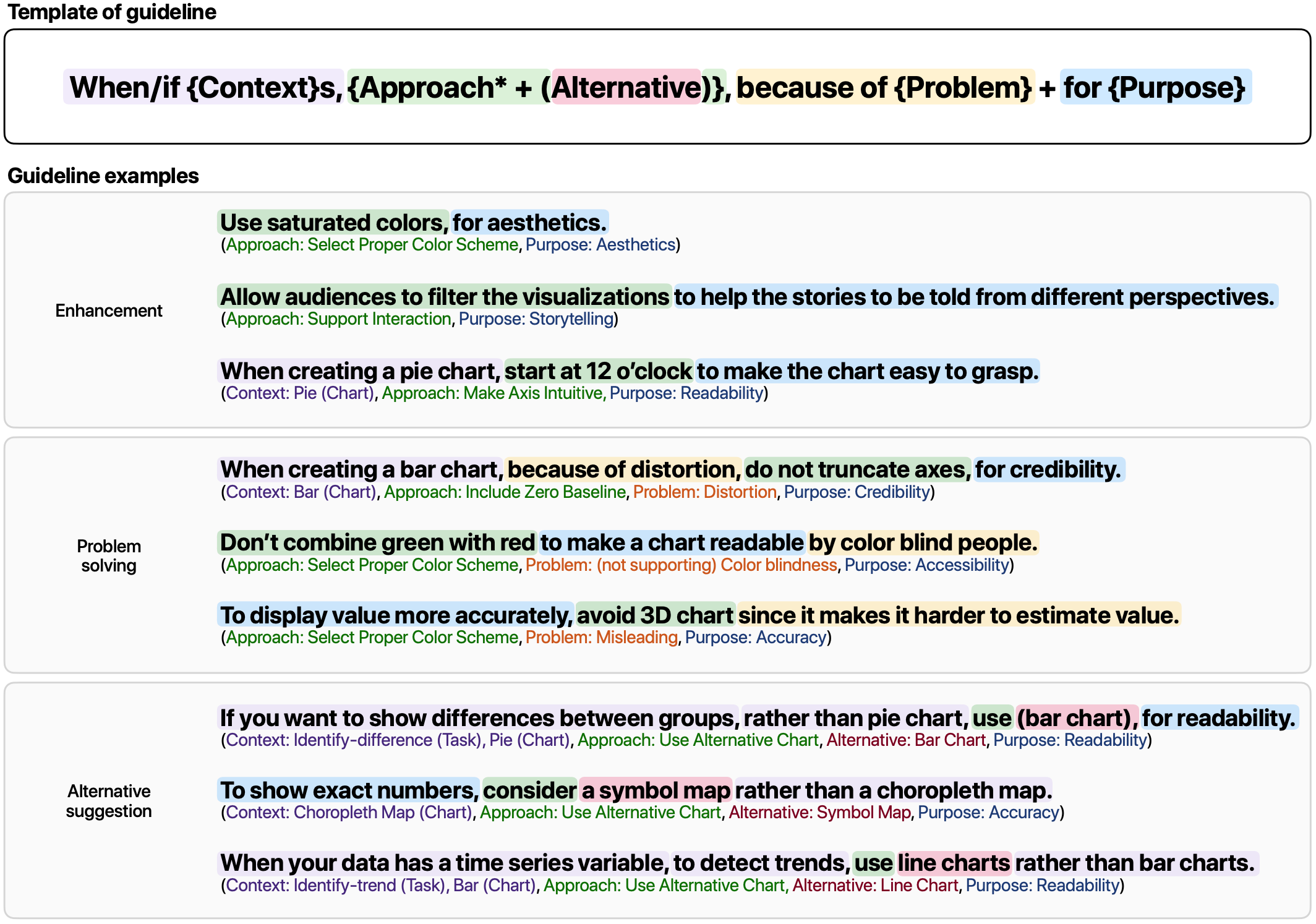}
    \caption{The structure template of guidelines and examples across the three different characteristics: enhancement, problem-solving, and alternative suggestion. The template consists of a combination of attributes and components of a sentence. Each component of the sentence is color-tagged according to each attribute it includes.}
    \label{fig:template}
\end{figure*}

While inspecting guideline sources during the coding process, we observed recurring patterns in the guideline structures. We categorized and developed them into three templates: \textit{problem-solving}, \textit{enhancement}, \textit{alternative suggestion}. We synthesized these patterns to derive a holistic guideline template (\autoref{fig:template}): \texttt{What/if \{\textit{Context}\}s, \{\textit{Approach}* + (\textit{Alternative})\}, because of \{\textit{Problem}\} + for \{\textit{Purpose}\}}. 

\textit{Problem-solving} guidelines include the ``problem'' attribute as a necessity, as they suggest to solve the problematic situation; e.g., \textit{distortion}, \textit{misleading patterns}, and \textit{clutter}. Most guidelines belong to this category (67\%), with common purposes including readability, accuracy, and clarity; for example, \textit{to accurately depict trends, ensure that the aspect ratio for trend charts is set to 45 degrees.}

\textit{Alternative suggestion} guidelines are similar to problem-solving guidelines, but they warrant separate consideration given their prevalence in practice. Most of the guidelines that have \textit{use alternative chart} (24\%) and \textit{add different encoding} (8\%) approaches belong to this category; for instance, \textit{when you need more than six categorical colors, try alternative visual encoding such as position}. 

\textit{Enhancement} guidelines do not necessarily have issues to be resolved. The primary objectives of these guidelines include \textit{storytelling}, \textit{engagement}, and \textit{aesthetics}, contributing to enhanced visualization quality and improved user experience. Example guidelines include: \textit{when formulating titles, use concise and active phrases that encapsulate the intended argument instead of descriptive titles}; and \textit{use graphics like icons only when they support interpreting the graph}.

The existing guideline description lacked a cohesive structure. These templates can provide a well-defined framework for effectively conveying actionable guidance to designers, akin to efforts in medicine for developing models of guideline representation~\cite{ohno1998guideline}. We used these templates to generate concise one-sentence descriptions for each guideline, further streamlining our open coding process.

\subsection{Guideline and Empirical Study Alignment}

\begin{figure*}[tb]
    \centering
    \includegraphics[width=\linewidth]{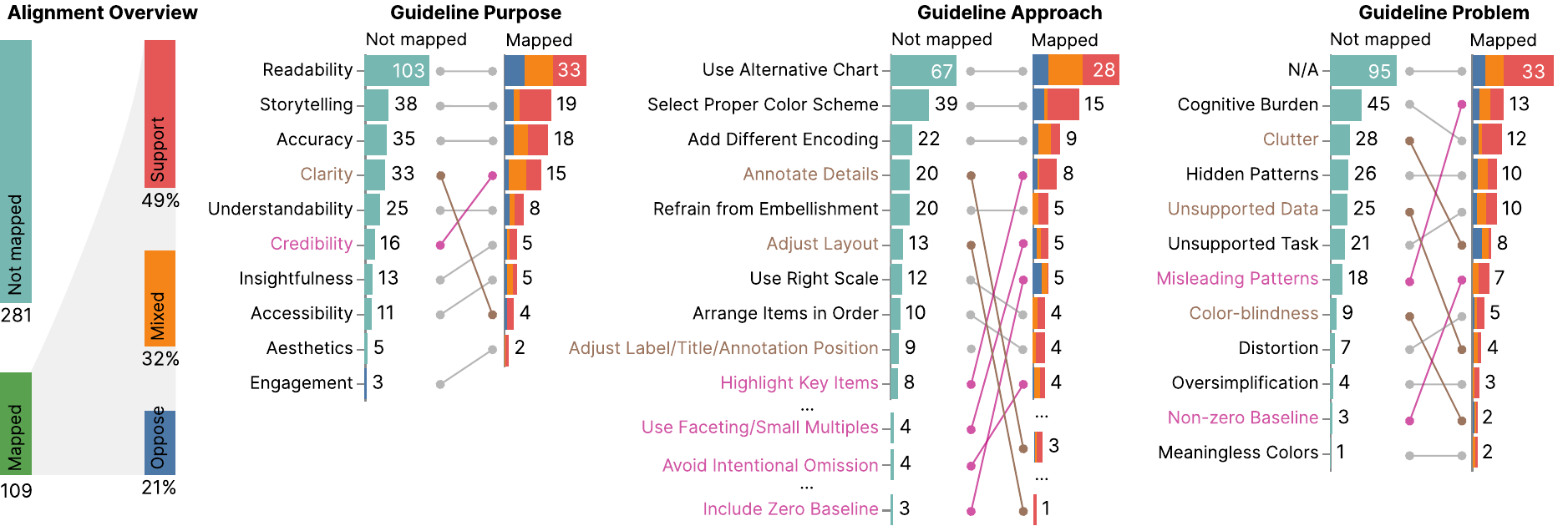}
    \caption{Distribution of Visualization guidelines mapped versus unmapped to empirical studies: This figure delineates the hierarchy of guideline purposes, approaches, and problems with notable shifts observed in rankings when mapped to empirical studies, indicating a tendency towards guidelines with empirical verification.}
    \label{fig:guideline-mapped}
\end{figure*}

Among the 390 collected guidelines, 109 (28\%) were mapped to empirical studies. Among the 235 empirical studies, 94 were linked to the guidelines. In other words, each guideline was often mapped to more than one empirical study ($avg$ / guideline = 2.00, $std$=1.62), while an empirical study was also mapped to more than one guideline ($avg$ / empirical study = 2.15, $std$=1.72). Many of the highly mapped guidelines and empirical studies were about well-known graphical perception issues such as regarding pie charts, embellishments, color maps, and scales \& axes. 

Among the total 219 guideline and empirical study mappings, the majority (49\%) were found to be \textit{supported} by corresponding empirical studies. However, we also observed \textit{mixed} support (32\%) or even \textit{conflicting} results (21\%) from empirical studies. For example, a guideline suggests avoiding the use of 3D charts because they can skew perception. However, a study found that 3D heatmaps were superior to 2D heatmaps for reading and comparing single data items, but 2D heatmaps were better for overview tasks~\cite{kraus2020assessing}. Similarly, some guidelines recommend minimalist chart designs without embellishments to avoid distracting from the data, but studies suggest that visual embellishments such as icons can improve long-term recall and interpretation accuracy~\cite{haroz2015isotype}. Similarly, other guidelines recommend avoiding rainbow colormaps, while several empirical studies contradict this guideline by demonstrating the value of multi-hue maps on continuous data for class discriminability~\cite{liu2018somewhere}.

We also looked at whether any different trends exist in the mapped and unmapped guidelines. The distribution of \textit{unmapped} guidelines mostly remained similar to the overall distribution mentioned in the guideline landscape analysis (Sec. \ref{sec:guideline-landscape}). However, there were several meaningful changes in the topic distribution of the \textit{mapped} guidelines (\autoref{fig:guideline-mapped}), indicating that empirical studies tend to prioritize verifiable guidelines over qualitative ones. For instance, we observed a notable drop in the frequency of the \textit{clarity} dimension within the guideline purpose category. Likewise, the frequency of the \textit{annotate details} and \textit{adjust layout} dimensions were much lower in the mapped guidelines, while some other approaches were not mapped at all, including \textit{adjust label/title/annotation position}, \textit{improve axis label format}, and \textit{choose the right orientation}. Similarly, \textit{misleading patterns} and \textit{non-zero baseline} were much more frequent in the mapped guidelines. For chart types and chart element types, bar, line, and pie charts consistently remained at the top, while scatter plots and axis elements (compared to annotation elements) more prominent in the mapped guidelines. We did not observe noticeable trends in task types.

\subsection{Takeaways}

The analysis of guidelines and their alignment with empirical studies has yielded valuable insights into design guidelines currently available and its empirical backing. We found that most guidelines primarily address basic charts, highlighting a gap in advanced visualizations like parallel coordinates and network diagrams. Interestingly, empirical evidence sometimes contradicted these guidelines, indicating the \textbf{presence of mixed and conflicting knowledge}. While variation is expected due to different contexts, it is potentially problematic when practitioners are \textbf{not made aware of these nuances}. Likewise, many guidelines lacked specific information on data and task types, critical for understanding the applicability of guidelines. Finally, the presence of unmapped empirical studies suggests opportunities for refining or creating new guidelines. Similarly, guidelines that remain unmapped point to areas where further empirical research is needed to develop solid evidence.

\subsection{Limitations}

While the core messages we intend to convey remain intact, the precise mapping statistics are transient, as they can easily shift with the emergence of new guidelines and the choices made regarding data sources and inclusion/exclusion criteria. As mentioned earlier, guidelines found in the wild are highly heterogeneous and difficult to categorize precisely due to the varied language, structure, and data quality used in different sources (e.g., the use of very generic language). We applied a reasonable rationale and perspective when selecting guidelines for our analysis (Section~\ref{sec:collecting-guidelines}), aiming to highlight meaningful gaps to address. However, we discourage draw any significant conclusions from the absolute or relative percentage differences. 

Moreover, our inclusion of less common guidelines may overstate the perceived knowledge gap, although this inclusion is important as it highlights underexplored areas of design knowledge. In addition, our alignment analysis was limited to empirical studies within the visualization literature, and it is possible that some unmatched guidelines could be supported by evidence from broader fields, such as cognitive psychology or art theory, which were outside the scope of this work.

\section{Practitioners and Researchers' Attitudes and Experiences}
\label{sec:surveys-interviews}
The content analysis allowed us to glimpse the current knowledge gap in guidelines and empirical research. However, it does not capture practitioners' and researchers' perceptions or their real-world experiences with the guidelines and the knowledge gap. To gain a deeper understanding and context, we conducted follow-up surveys and interviews with stakeholders from both groups.

\subsection{Survey Recruitment \& Procedure}
Our recruitment methodology varied slightly between practitioners and researchers. For practitioners, our recruitment targeted individuals engaged in data visualization, regardless of their specific job titles and responsibilities, similar to prior work in data visualization practice~\cite{parsons2021understanding}.  This recruitment was based on the premise that knowledge of data visualization design is essential for anyone who creates visualizations. 
We reached out to members of the Data Visualization Society's Slack workspace~\cite{datavissociety}, focusing on individuals who had shown recent engagement through activities such as posting messages in public channels. Our approach involved the use of direct messaging, intended to improve the overall participation rate. Additionally, we endeavored to expand our reach through postings on the Storytelling with Data forum~\cite{storytellingwithdata} and direct messages to visualization practitioners on Twitter, although these efforts were not as successful as the initial approach.

For the recruitment of researchers, we pursued an alternative approach, reaching out to key stakeholders via email. Specifically, we contacted individuals associated with the \href{mailto:infovis@infovis.org}{infovis@infovis.org} and \href{mailto:ieee\_vis@listserv.uni-tuebingen.de}{ieee\_vis@listserv.uni-tuebingen.de} mailing lists, as well as researchers with expertise in empirical studies in visualization. Our targeted population included esteemed faculty members, post-doctoral researchers, and Ph.D. students in candidacy, all of whom were deemed suitable candidates for our research endeavor.

We distributed the surveys online, and participants took approximately 6-10 minutes to complete each. The surveys contained similar questions, categorized into sections on background, awareness, opinions on visualization design guidelines, and perceptions and expectations of empirical research in visualization (see example questions in \autoref{fig:survey-guidelines} and \autoref{fig:survey-empirical-studies}). 
We presented participants with examples of misleading visuals from WTF Viz~\cite{wtfviz} to gauge their awareness of such charts. We also included a screenshot of a comparative experiment between bar and pie charts to assess familiarity with empirical research. The full survey questionnaires can be found in the supplementary material. To compensate participants, we offered a \$25 raffle reward for every 25 participants who completed the survey.

\subsection{Survey Participants}

Overall, 33 practitioners and 36 researchers filled out the surveys. Practitioners indicated their jobs as the following: eight analysts (24\%), six designers (18\%), six scientists (18\%), three managers (9\%), three developers (9\%), two teachers (6\%), one journalist (3\%), and four others (12\%). The others category includes one project lead, one consultant, and two engineers. Sixteen of them (48\%) had 3-5 years of experience, eight had 11-15 years (24\%), six had 6-10 years (6\%), two had 0-2 years (6\%), and one had more than 16 years of experience (3\%). In terms of research areas, 20 researcher participants indicated applications (24\%), followed by 19 indicating empirical research (23\%), 16 indicating representation \& interaction (19\%), 10 indicating theoretical work (12\%), eight indicating analytics \& decisions (10\%), seven indicating systems \& rendering (8\%), and three indicating data transformation (4\%).

\subsection{Survey Results}

\begin{figure*}[tb]
  \centering 
  \includegraphics[width=\linewidth]{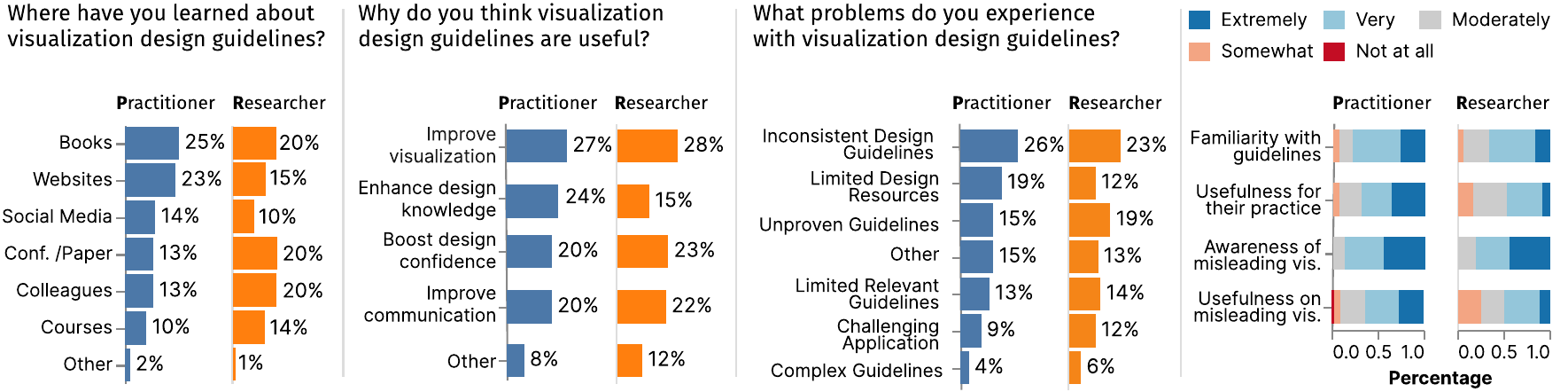}
  \caption{Responses to a survey on design guidelines, showing sources of learning, perceived usefulness, and experienced problems. Books and websites were the most common learning sources, while improving visualization knowledge was the top benefit noted. Inconsistent guidelines were the main issue faced. Degrees of familiarity, application to practice, awareness of misleading visuals, and overall usefulness were also reported.}
  \label{fig:survey-guidelines}
\end{figure*}

\begin{figure}[tb]
  \centering 
  \includegraphics[width=\linewidth]{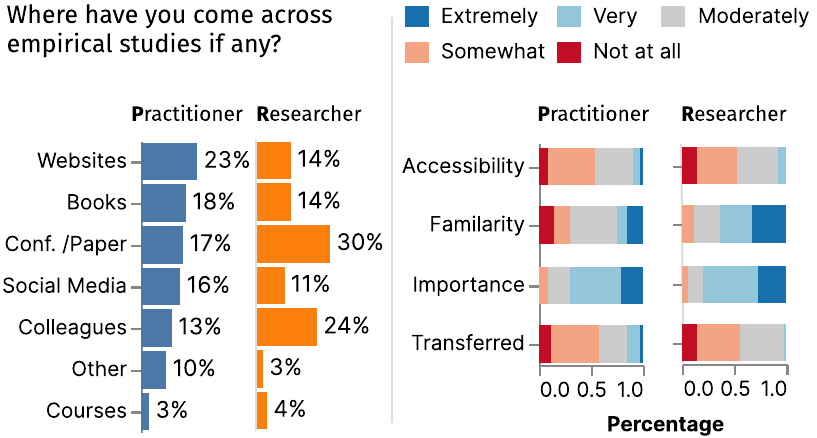}
  \caption{Survey findings on the discovery of empirical studies, with participants predominantly encountering them through websites and conferences. The graph also reflects levels of accessibility, familiarity, perceived importance, and the extent to which insights are transferred into practice.}
  \label{fig:survey-empirical-studies}
\end{figure}

\autoref{fig:survey-guidelines} and \ref{fig:survey-empirical-studies} provide an overview of survey responses. Generally, practitioners and researchers showed similar perceptions and opinions toward design guidelines and empirical studies, but differences were observed in their backgrounds and experiences. For example, practitioners reported that they learned about guidelines mainly from books (25\%) and websites (23\%), such as online articles and blogs. In contrast, researchers relied more on research papers (20\%) and colleagues (20\%) to learn about guidelines, with a relatively higher percentage compared to practitioners (13\% for each category). Regarding empirical studies, practitioners indicated that their primary access channels were online resources, such as websites, articles, and blogs, accounting for 23\%. In contrast, researchers indicated conferences and research paper archives the most, representing 30\% of their preferred access channels.

Both stakeholders agreed that guidelines could be beneficial for creating better visualizations (\textbf{P}ractitioners: 27\%, \textbf{R}esearchers: 28\%),  improving confidence in design decisions (P: 20\%, R: 23\%), and facilitating better communication with stakeholders (P: 20\%, R: 22\%). However, researchers weighed the usefulness of guidelines on increasing visualization design knowledge less (15\%) than practitioners (24\%). In the \textit{Other} comment section, several practitioners mentioned additional benefits of guidelines: serving as a \textit{``standard of practice''}; helping avoid \textit{``misleading, ambiguous''}, and \textit{``deceitful''} visualization designs; increasing \textit{``user understanding''} and \textit{``trust''}, producing \textit{``consistent user experience''}; \textit{``facilitating design decision making''}. Researchers' comments were similar, while they additionally mentioned that guidelines provide a \textit{``scaffold''} or \textit{``starting point''} for learners.

When asked about the problems they face with existing guidelines, both stakeholders identified \textbf{inconsistent and conflicting guidelines} as the most frequent problem (P: 26\%, R: 23\%). Practitioners weighted more on the \textbf{lack of design resources} (P: 19\%, R: 12\%), while researchers emphasized more on the \textbf{trustfulness of guidelines} (P: 15\%, R: 19\%). In the \textit{Other} comment section, practitioners provided additional comments:\textit{``[a guideline] limits the ability to take into account semantics and intent''}; \textit{``[it is] difficult to explain to stakeholders why they should follow those guidelines''}; \textit{``[it's] best to use own judgment. When it comes to bad design, it's usually bad intentions getting in the way''}.
The comments of the researchers echoed similar sentiments: \textit{``better designs depend critically on context, and guidelines are context-free''}; 
\textit{``sometimes [they're] overgeneralized, leading to unwarranted confidence in design decisions''}; \textit{``[it is] not always easy to find authoritative sources for guidelines, which makes it challenging to convince stakeholders/clients''}.

Both practitioners and researchers indicated that they are generally familiar with design guidelines (\autoref{fig:survey-guidelines} rightmost). As expected, practitioners found design guidelines more relevant to their practice. Although both groups generally understood the concept of misleading visualization designs, practitioners placed a greater emphasis on the usefulness of guidelines for addressing such issues (\autoref{fig:survey-guidelines} rightmost), compared to researchers who did not prioritize guidelines to the same extent as practitioners. When we asked researchers how they communicate their research to the public, if any, most said: conference presentations (41\%), followed by social media (26\%) and online resources such as blogs (20\%).

When we asked about the availability of empirical studies, both practitioners and researchers agreed that empirical studies are generally not very accessible to practitioners (\autoref{fig:survey-empirical-studies}).  Despite differences in familiarity with empirical studies, there was a shared recognition of the significant role empirical studies play in producing design guidelines (\autoref{fig:survey-empirical-studies}). Similarly, they agreed that the current transfer of knowledge from empirical research findings into practical visualization guidelines might not be effective (\autoref{fig:survey-empirical-studies}).

When we generally asked for additional comments on guidelines, several practitioners commented that it would be useful to \textit{``update examples''} of guidelines, while it is hard to find which guidelines are \textit{``relevant''} and \textit{``trustworthy''}.
On the other hand, researchers suggested \textit{``including information about their limitations,''} avoiding the tendency to draw \textit{``a false dichotomy between good and bad,''} and taking into account \textit{``practical differences''} when making recommendations between one approach and another.

Practitioners did not provide extensive comments on empirical studies, although one participant emphasized the importance of authors  \textit{``translating empirical work into practical applications''} for practitioners who may not have the time or inclination to read academic papers. On the other hand, researchers indicated the tendency to \textit{``prioritize efficiency''} excessively, limited \textit{``meaningful outreach with practitioners''}, and the \textit{``overlooking of practical contexts''} in empirical studies.


\subsection{Interview Procedure}

Two researchers conducted interviews remotely. The interview questions were prepared in advance and tailored to each interviewee's survey response to understand their experiences and perspectives in depth. For instance, if participants selected a particular category of challenges, we probed further by asking them to elaborate on their perceptions and ideas. The questions were structured to follow the same format as the survey questionnaire and were kept consistent across all practitioners and researchers involved in the interviews. One researcher facilitated the questioning process, while the other researcher took notes and transcribed the interviews. The complete list of interview questions for both practitioners and researchers is available in the supplementary material. Interviewees received a compensation of \$50 for their participation.

\subsection{Interview Participants}

\begin{table*}[]
\centering
\begin{tabular}{llll|llll}
\hline
\multicolumn{4}{l|}{Practitioners Interviewees}             & \multicolumn{4}{l}{Researcher Interviewees}                              \\ \hline
ID  & Job Title           & Experience & Guideline Familiarity               & ID  & Job Title                   & Experience     & Guideline Familiarity                \\ \hline
P1  & Manager             & 6-10 yrs   & Extremely familiar & R1  & Postdoc researcher     & 11-15 yrs      & Very familiar       \\
P2  & Product Lead        & 3-5 yrs    & Very familiar      & R2  & Graduate assistant & 0-2 yrs        & Moderately familiar \\
P3  & Scientist           & 6-10 yrs   & Very familiar      & R3  & Graduate assistant & 6-10 yrs       & Extremely familiar  \\
P4  & Consultant & 11-15 yrs  & Extremely familiar & R4  & Postdoc researcher     & 6-10 yrs       & Very familiar       \\
P5  & Scientist           & 3-5 yrs    & Very familiar      & R5  & Research scientist          & 3-5 yrs        & Extremely familiar  \\
P6  & Developer           & 6-10 yrs   & Extremely familiar & R6  & Graduate assistant & 3-5 yrs        & Moderately familiar \\
P7  & Manager             & 11-15 yrs  & Extremely familiar & R7  & Faculty                     & 11-15 yrs      & Very familiar       \\
P8  & Analyst             & 3-5 yrs    & Very familiar      & R8  & Faculty                     & 16 or more yrs & Slightly familiar   \\
P9  & Designer            & 3-5 yrs    & Extremely familiar & R9  & Faculty                     & 16 or more yrs & Very familiar       \\
P10 & Analyst             & 11-15 yrs  & Extremely familiar & R10 & Research scientist          & 16 or more yrs & Very familiar       \\ \hline
\end{tabular}
\caption{Demographic characteristics of practitioner and researcher interviewees}
\label{tbl:interview-participants}
\end{table*}

After following up with survey participants to inquire about their availability for interviews, 21 out of 33 practitioners agreed to participate. Thus, we selected ten participants based on differing opinions on how well knowledge of empirical research transferred to visualization practice (one thought it transferred extremely well, two thought it transferred very well, two thought it transferred moderately well, three thought it transferred slightly well, and two thought it did not transfer well at all). Upon choosing the participants, we also considered participants who are more acquainted with design guidelines (six were extremely familiar, and four were very familiar), as they can provide more substantial perspectives on the knowledge gap associated with the guidelines.

Out of the 36 researchers, 16 expressed interest in participating in the interviews. We carefully chose ten interviewees who have familiarity with conducting empirical experiments (three participants were extremely familiar, five were very familiar, and two were slightly familiar) as their insights can help uncover nuances in the challenges and opportunities associated with the current knowledge transfer effort. The selected participants exhibited varying levels of familiarity with visualization design guidelines (two were extremely familiar, five were very familiar, two were moderately familiar, and one was slightly familiar), which allowed us to capture a broad spectrum of viewpoints on how design guidelines are perceived and integrated into empirical research practices.

\subsection{Interview Analysis}
All interviews were conducted within a 30 to 45-minute timeframe, and each was transcribed using an automatic transcription service. The transcripts were then manually checked for errors. During the initial coding process, a research assistant pulled related quotes from the transcripts, excluding unrelated conversational text for efficient analysis. The assistant coded them in a spreadsheet using the interview questions as the initial set of codes. The lead researcher reviewed and refined the codes. They repeated this process for approximately one-third of the interviews until arriving at a consistent set of codes (i.e., inductive coding) . The research assistant then applied the final codes to the remaining interviews, which were comprehensively reviewed by the lead researcher (i.e., deductive coding). Overall, we categorized the quotes into eleven themes: five focused on experience and perception, challenges related to guidelines, two themes concerning perception and accessibility of empirical studies, four themes addressing the perception of the knowledge gap, strategies to mitigate this gap, and expectations within the visualization communities.


\subsection{Interview Results: Guidelines}

\subsubsection{Personal Definitions of Visualization Guidelines}
 Practitioners viewed visualization design guidelines as a set of best practices, principles, or rules that guide the creation of effective and consistent visualizations. They noted that guidelines help organizations improve the way they communicate data and to be able to make better decisions, covering things like axis labeling and composition strategies (P1, P2, P6, P9). However, they also highlighted that there are not any absolute rules in data visualization and that it is important to \textit{``tailor what you do and tailor what you make to your audience and to your data''} (P6). 
 
 Researchers generally agreed that guidelines are a set of basic rules of good or bad practice.
Although practitioners frequently viewed guidelines as more prescriptive (e.g., maintaining consistency, brand identity), researchers viewed them as more flexible heuristics that can adapt to different situations and audiences. For instance, R2 shared \textit{``it's just the designer's intuition on what looks right and what doesn't.''}, while R3 commented, \textit{``each organization or designer [...] usually come[s] up with their sets of soft guidelines versus the hard ones.''} They also emphasized that guidelines are not meant to be \textit{``authoritative''} (R6) and current guidelines are often \textit{``written more prescriptive than they probably should be''} (R10).


\subsubsection{Examples of Visualization Guidelines}

The examples provided aligned with our previous content analysis findings. These examples include zero baselines (P3, P6), issues with pie charts (P8, P9), minimizing color use (P2, P4), and color blindness (P8). For instance,  P2 commented, \textit{``HDX guideline [recommends to] minimize the amount of color.''}, while P8 similarly noted \textit{``[...] accommodate other disabilities, color blindness as much as possible.''} P3 also shared the commonly known guideline, saying \textit{``if you have a bar chart you shouldn't cut the y axis.''} Practitioners also commented on broader issues such as grid lines and sizing (P1), layouts \& organizations (P1, P2), visual attention (P7, P10), visual branding (P7), and storytelling (P8). Other practitioners shared some exceptions to the guidelines such as using donut charts instead of pie charts (P8) and using non-zero baselines (P6---\textit{``I don't necessarily think that always has to be the case as long as you're upfront about it''}).

Researchers echoed practitioners in some areas, reiterating guidelines on zero-baselines (R1, R10) and pie charts (R7). But they also shared more specific design guideline knowledge such as rainbow color maps (R2, R6, R9), length vs angle vs area (R1, R2, R10), comparison along a common-scale vs unaligned-scale (R1, R8), 3D charts (R4), and banking to 45 degrees (R9). Some of these guidelines were neither found on our guideline analysis nor discussed by practitioners. Researchers were often critical in their discussions, noting exceptions to general guidelines and pointing out vagueness in some common advice. For instance,  R6 said, \textit{``there are certain cases when with the scientific data it makes sense to actually utilize [rainbow colormaps]''}, while R7 said, \textit{``One vis design guideline that we hear a lot about is don't use pie charts, which I would take to be a bit vague.''}

\subsubsection{Lived Experiences with Visualization Guidelines}

Practitioners had mixed views on the applicability and frequency of using visualization design guidelines in their practice. They mentioned that they do not strictly adhere to the guidelines but rather underscored the significance of taking into account user needs and feedback (P1, P4, P5, P7). 
They also pointed out that experience, intuition, and learning from others played crucial roles in shaping their decision-making process (P5, P6, P7, P9, P10). For instance, as their experience grows, they rely less on guidelines (P6) and instead trust their instincts and inspirations (P5, P9). Similarly, P10 noted that observing real-world applications in projects helps them gauge the utility of a guideline. Despite this somewhat critical perspective on guidelines, other practitioners emphasized that guidelines provide fundamental knowledge and foster critical design thinking (P5, P9).


The responses from researchers were generally consistent with those of practitioners, reinforcing the notion that there is no universal solution for visualization design guidelines. They also underscored the significance of considering the project's context, objectives, audience, and collaborators. For instance, R3 noted that guidelines are \textit{``flexible tools that can accommodate the needs and viewpoints of various stakeholders involved in visualization design.''} On the other hand, R10 emphasized that guidelines are valuable to navigate the fluidity and vagueness inherent in data visualization design, particularly for beginners, even though \textit{``they are not so clear cut and should be taken with a grain of salt.''}





\subsubsection{How They Encounter Visualization Guidelines}

When inquiring about their sources for accessing data visualization guidelines, practitioners mentioned a blend of formal and informal channels, including books, websites, podcasts, and professional networks. For instance, they referred to resources offered by notable companies and organizations like Google, Apple, IBM, and HDX (P1, P2). On the contrary, P4 discussed a collection of books authored by Tufte, Few, and Knaflic that predominantly focus on charts, contrasting them with the book by Tamara Munzner, which offers more theoretical models and knowledge. Others cited specific resources, such as the Data Stories podcast (P3), Medium blogs (P7), Twitter posts (P6), and additional resources such as Nightingale Magazine available through the Data Visualization Society (P2, P6, P7, P10). In general, they do not actively seek out data visualization design guidelines. Instead, they tend to encounter guidelines organically, such as when researching aspects of their projects (P9). However, they do deliberately select particular books to read, especially those focused on topics like dashboard design (P10).


In general, researchers encountered visualizations in a similar fashion (e.g., books, blogs, social media, etc.), albeit with a more frequent mention of academic sources. They specifically noted encountering guidelines through papers presented at conferences like ACM CHI and IEEE VIS (R1, R6, P7 R8). Similarly, R2 described \textit{``absorbing''} guidelines \textit{``passively through reading papers,''} while others referred to course materials (R1, R4, R10) or academic workshops (R9). Interestingly, R7 highlighted that effective visualizations crafted by creative professionals or journalists often act as valuable design guidelines.




\subsubsection{Perceived Usefulness of Visualization Guidelines}

Practitioners had generally positive opinions regarding the usefulness of visualization design guidelines. They expressed that they found them helpful because they provided a framework and direction for visualizing data, simplified their work, and made their communication more effective (P3, P5, P7, P9, P10). For instance, P9 said, guidelines \textit{``help me communicate what I want to do to other people [...] also help people to understand what I'm saying.''}, and further elaborated that \textit{``it's like basic knowledge you have to have [...] to apply for your job.''} However, P1 indicated less enthusiasm toward guidelines due to their lack of applicability and thoroughness.



Researchers shared positive perceptions akin to practitioners. R1 pointed out that guidelines can help \textit{``convince coworkers''} and themselves toward better designs. Others indicated that they use guidelines \textit{``mainly to teach''} students (R6) and to help with the \textit{``exploration of the design space''} of what works and what does not (R7). In addition, R2 noted that \textit{``they are useful in the sense that if they are integrated into some kind of system, so like Tableau, for instance, automatically select[ing] correct color scales.''} Nevertheless, researchers also highlighted that guidelines are less effective without contextual examples (R7) and often lack sufficient actionability unless accompanied by clear alternative suggestions, even if the conceptual foundations behind the guidelines hold value (R9).




\subsubsection{Struggles and Challenges with Visualization Guidelines}

Practitioners observed that the existing guidelines have limitations and do not adequately cover emerging concerns like interactivity and narrative elements (P1, P4). They also emphasized that the present guidelines seem overly constrictive, lacking an embrace of subjectivity and the contextual nuances of diverse projects (P4, P6, P7, P9). For instance, P6 said, \textit{``with climate change [...] small differences have a really big effect [...] always including like zero degrees in the y-axis just doesn't make any sense.''} Similarly, a number of practitioners highlighted encountering trade-offs with business requirements (P3, P8, P9). For example, as noted by P3, \textit{``if [..] they already plotted their data [...] a certain way [...], they might not want to change it.''} Likewise, they indicated that effectively disseminating the guidelines to accommodate a variety of stakeholders (designers, analysts, managers, etc.) presents a distinct challenge (P2, P3, P4, P10). What exacerbates the situation is the lack of consensus in the field regarding what constitutes good design, resulting in conflicting and scattered guidelines (P2, P3, P4, P7, P10). Hence, they also emphasized the necessity for improved examples of how to apply the guidelines and when to deviate based on specific situations, rather than solely relying on negative instances that advise against certain practices (P5, P7, P8, P9, P10).




Researchers were mainly concerned about challenges in making guidelines more accessible and useful for practitioners. One common theme was the difficulty of aligning the guidelines to specific user requirements and design goals (R1, R2, R3, R8, R10), as R2 noted, \textit{``I couldn't really type in like I have this and what should I use?''} They also underscored the overwhelming number of scattered guidelines, along with the challenge of staying abreast of new research (R2, R8, R9). For instance, R10 shared \textit{``there are so many tasks, so many types of data and so many variations to think...''}, while R9 pointed out the \textit{``impossibility of actually making sense of all the studies out there.''} As with practitioners, many researchers expressed frustration with the conservative nature of existing guidelines and their limited versatility, suggesting the need to acknowledge contextual nuances and offer practical flexibility (R4, R7, R8, R9, R10). In a similar vein, they highlighted the necessity for clearer communication of provenance and limitations (R7), addressing conflicting arguments (R6), and establishing a shared vocabulary across disciplines to facilitate the interpretation of guidelines (R5). Moreover, the overarching lack of consensus on design standards and the variation in outcomes from empirical studies seem to further complicate the establishment of universally applicable guidelines (R9, R10).









\subsection{Interview Results: Empirical Studies}

\subsubsection{Perceived Usefulness of Empirical Studies}

All practitioners generally agree that empirical studies provide evidence and validation for design choices. Several practitioners stressed the value of helping to better understand the reasoning behind graphical selections (P2, P4, P5, P6, P7, P10). For instance, P10 commented, \textit{``it's really helpful for me to see the research that says, if you're in this situation you can do this or like in this situation [...] this is how the participants interpreted this graph.''} P6 appreciated anecdotal evidence that had been rigorously validated, while P5 highlighted that it saves them time by sparing them from the need for trial and error. Others said empirical research helps to \textit{``convince those naysayers and the extreme skeptics among us''} (P6), by \textit{``having proof''} (P5). On the other hand, P1 emphasized the importance of qualitative research too, saying \textit{``doing research like how people feel when they look at visualization is important as well.''}

Although researchers held similar opinions, some expressed skepticism about the practicality of using empirical results in design practice. For instance, R1 noted, \textit{``design guidelines were from refining the results of empirical studies, so I think there are some differences between the two.''} Likewise, others highlighted the need for qualitative research to gain a holistic understanding of user preferences and behaviors (R2, R10), as R2 said \textit{``if you just use [...] empirical quantitative stuff it might not always capture that enjoyment and stuff aspect of visualization.''} Several interviewees stressed the importance of replicating studies and verifying the statistical significance of claims. For instance, R6 said, \textit{``I'm not sure if I ran in one place and I ran with a different group of people, we'll get the same results''}, while R9 commented, \textit{``there are different levels of evidence behind guidelines.''} In addition, R7 emphasized empirical research should be a concerted effort involving multiple parties, including \textit{``researchers, designers, and tool developers.''}



\subsubsection{Accessibility of Empirical Study Findings}
Participants acknowledged challenges in accessing empirical research, citing factors such as paywalls (P2, P3, P6), complex language (P3, P4), and time constraints (P6). For instance, P4 noted \textit{``the language of academia is not fully accessible to practitioners.''} P6 similarly added, \textit{``The amount of time to find, read and understand is a big barrier for entry.''} They utilize a variety of sources to access empirical studies, academic websites, journals, blogs, and even networking platforms like Twitter (P2, P4, P5, P8, P10). Several practitioners expressed a preference for summarized findings and design guidelines (P6, P7, P10). For instance, P7 noted, \textit{``I would want to know what the results are and how it impacts my work.''}, while P6 said, \textit{``I think for most people ... guidelines are good enough.''}






Researchers expressed difficulty in accessing empirical studies for practitioners, mainly due to complex jargon and limited availability of easy-to-understand resources (R1, R5, R6, R8, R10). They also acknowledged that paywalls can hinder practitioners' access to valuable research (R6, R10), as well as the lack of necessary background knowledge (R8). Some participants suggested that providing access to well-organized and summarized research could improve accessibility (R1, R10). On the flip side, some said practitioners might not need to check studies themselves (R2, R5), while others pointed out a need to understand the reasons behind guidelines (R6, R8, R9).








\subsection{Interview Results: Knowledge Gap}

\subsubsection{Perception and Attitude toward Knowledge Gap}

Practitioners expressed differing opinions on the existence of a gap between research findings and practical knowledge. Some noted a significant gap, where information from conferences takes time to reach public books (P3) or mentioned challenges in translating research into practical applications (P4). Others felt there was no gap due to the availability of information if one knows where to look (P8). On the other hand, some others were unsure about the extent of the gap since not all practitioners might engage with research directly, possibly contributing to the perceived gap (P9, P10).





Researchers also indicated varying perceptions of the knowledge gap. Some of them emphasized the challenges in accessing and understanding research studies for practitioners (R1, R4, R5, R6); for instance, R5 attributed the issue to the interdisciplinary nature of the field, while R6 pointed out the contradicting knowledge spread across the field. Others highlighted that individuals who publish guidelines in blog posts or company-driven guidelines might lack the necessary background to effectively convey knowledge. For instance, as mentioned by participant R7, \textit{``it pays to have a blog [...] because it increases your exposure on the web.''} On the other hand, R2 believed that the gap is not significant, saying \textit{``Use like a scatter plot if you have two variables. Bar charts work great for 90 percent of things. I don't know how many more design guidelines we need.''}





\subsubsection{Expectations for Visualization Communities}
The practitioners expressed a need for better communication and collaboration between researchers and practitioners. For instance, P4 highlighted the necessity of terminology adaptation between practitioners and researchers. Meanwhile, P1 emphasized the importance of practitioners being involved in research to make the outcome more actionable. They also emphasized the importance of making research more accessible and applicable to practitioners, including real-world applications of the studies (P7), broader access to facilitate the implementation of findings (P8), and creating guides and summaries (P5, P10). Some expressed a degree of cynicism about the possibility of successful collaboration (P6), while others remained hopeful for more open data and knowledge sharing (P9).

The researchers equally emphasized the importance of effective communication and collaboration between the two groups (R3, R5, R6). R6 specifically mentioned that researchers should take extra steps to reach practitioners, such as creating casual articles, videos, and podcasts to make their work more accessible to practitioners. On the other hand, they said practitioners also have a responsibility to be critical of their sources and seek credible information, as R6 noted \textit{``Like you shouldn't believe every medical advice that's out there on the internet.''} R10 emphasized that \textit{``there probably should be more cross transfer between more practitioner-oriented conferences and more research-oriented conferences.''} Similarly, R7 conveyed a wish for the ongoing publication of guidelines, especially by practitioners, even in the face of potential adverse consequences, envisioning a growing space where both scientific and practitioner-oriented research coexist.

\subsubsection{Perceived Causes for the Gap}

Not a lot of practitioners articulate specific causes for the knowledge gap, except P4 who discussed the cultural and educational differences between practitioners and researchers, which can create tension and challenges in communication.
On the other hand, researchers pointed out the challenges in applying research findings universally (R1, R9); e.g., R1 noted, \textit{``the research was done in specific data, specific tasks, and in a particular environment, [so] it isn't easy to apply it universally.''} Some mentioned the limited interaction between practitioners and researchers poses a barrier to knowledge dissemination (R2, R3), while others highlighted the absence of research that caters to practitioners' specific needs and preferences (R4, R5, R9). They also noted that academic researchers might not usually bear the responsibility of disseminating their knowledge beyond their specific disciplines (R5, R10).  Several researchers also voiced criticism regarding the current reviewing system, which places more emphasis on new studies rather than replication and guideline development (R10). Additionally, they highlighted the absence of funding to support such endeavors (R9). 





\subsubsection{Ideas and Methods for Alleviating the Gap}

The main theme from practitioners was again centered around making resources more accessible and applicable. 
They suggested simplifying technical language to partnering with practitioners for applied studies (P4, P6, P7, P10), as P6 said, \textit{``It would be good if the academics would venture a little more out into practice. And it would be good if practitioners would open their minds a little bit''.} They also emphasized the importance of showcasing real-world examples (P1, P4, P6, P7), as well as simplified summaries of research findings (P10). For instance, P1 suggested augmenting existing programming libraries with design guidelines. Practitioners also advocated for publicly available content, such as workshops, newsletters, and courses, to enhance knowledge transfer (P2, P3, P5). In addition, they recommended increasing the visibility and accessibility of research through online platforms and open-access knowledge channels (P9).

Researchers' ideas were not very different but provided different angles for tackling the gap. Participants emphasized the importance of mutual understanding through workshops, collaborations, and sharing perspectives (R1, R6, R7, R8, R9). They expressed a need for practitioners' viewpoints in the design process and guidelines development (R1, R4, R6). Suggestions included making research findings more accessible through mediums like articles, videos, and blog posts (R1, R2, R6, R7). Furthermore, the significance of actionable design guidelines grounded in empirical evidence was highlighted (R3, R9). For instance, R9 elaborated, \textit{``And so having the pros and cons, having examples of where this went wrong, having examples of where actually it made sense to use a pie chart or what not yeah. So to give the practitioner a bit of a feel for what these arguments are about and what the layer of the land is there in the current research.''} Researchers also commented on incentives to engage in outreach and communication efforts (R7, R10). Finally, understanding the current knowledge transfer process and its potential harms \& incentives was highlighted as a crucial first step (R7).




\section{Takeaways}

Overall, there was a fair degree of agreement between both stakeholders, with slight differences in emphasis. They agreed that guidelines can be valuable for providing knowledge and educational benefits; however, the current formats are not ideal. Specifically, guidelines are often presented in a prescriptive manner and fail to consider many experiential and contextual factors. Similarly, both groups acknowledged the usefulness of empirical research but expressed concerns about its lack of accessibility in practice. It was evident that research papers, in their current form, are not suitable for practitioners, indicating a need for translation efforts, such as summaries or example-based descriptions using both good and bad examples.

Despite these points of agreement, there were also differences between the two groups. Practitioners rarely refer to research papers, and the guidelines they mentioned, as well as how they accessed them, differed from those cited by researchers. This validates our decision to filter research papers in the guideline collection and corroborates the discrepancies identified in Section 3. Regarding knowledge gaps, researchers were more aware of these gaps, while practitioners appeared less conscious of them, likely due to being limited by the resources available to them. In addressing these gaps, more effective engagement, such as providing publicly accessible content and fostering collaborative efforts (e.g., workshops, courses), were suggested, though there was some skepticism about the feasibility due to potential challenges, such as a lack of incentives.

\section{Limitations}
This study has several limitations. Sampling bias may be present, as participants were recruited from specific communities, which might not fully capture the diversity of practitioners and researchers. The qualitative nature introduces subjectivity in interpretation, despite efforts to ensure consistent coding. Additionally, self-reporting bias could impact our findings, as participants may have presented themselves more favorably on topics such as knowledge gaps or their use of guidelines. Finally, the varied experiences of participants make it challenging to aggregate their opinions into a unified perspective, which may affect the consistency of the findings.

\section{Discussion}

In this section, we reflect on a comprehensive strategy that integrates insights gained to enhance knowledge exchange between visualization research and practice.

First of all, there appears to be a need to \textbf{establish consistent vocabularies for describing data visualizations}, including chart names and task types. This issue was observed during our examination of guidelines and was reiterated in interviews with practitioners; e.g., there is often a lack of clear distinction between task types, chart types, and data types in the categorization of visualizations~\cite{datatoviz,visvocab,datavizproject}. While researchers may require a more granular vocabulary, such as the task taxonomy proposed by Brehmer and Munzner~\cite{brehmer2013multi}, it must be tailored to the level of practical needs and contexts. Previous research has similarly advocated for eliminating ambiguity in vocabulary related to chart junk~\cite{akbaba2021manifesto}, while others have argued that ambiguity in both research and practice can have its own value~\cite{shukla2022negotiating}.

A critical issue we observed is inconsistent design guidelines available in practice, as well as authoritative guidelines that do not acknowledge contextual factors. Interviewees often recognized this issue and suggested the need for \textbf{improving current design guidelines with empirical evidence details}. Empirical studies often contain mixed and conflicting findings, reflecting the diverse contexts in which these studies are conducted. However, the challenge lies in effectively communicating these complex nuances to practitioners. 

We can imagine having a convenient knowledge base where people can access design guidelines backed up with related empirical studies, as we did in our alignment analysis. Empirical evidence should be presented in language that is understandable to practitioners, and it should explain the context of the research so that they \textbf{understand its limitations and applicability}. This will improve the trustworthiness of each guideline and increase its adoption by practitioners.

We also observed how practitioners appreciate intuition and experience in design practice. We need to consider how to incorporate such experiential knowledge. VisGuides~\cite{diehl2018visguides} is an example of a platform where practitioners can share their examples for discussion and feedback. We can also imagine \textbf{connecting existing examples to guidelines} to create a \textbf{triadic model} of guidelines, empirical evidence, and relevant examples. The examples can demonstrate how people applied or deviated from the guidelines and present their success and failure cases. Furthermore, these examples might inspire researchers to develop future practice-oriented studies, thereby building a feedback and feedforward loop.

Participants also shared the difficulty of finding relevant guidelines. Therefore, \textbf{annotating the guidelines with detailed metadata}, such as tasks, chart types, or other contextual factors, would be necessary to facilitate efficient navigation of the landscape of visualization design guidelines. Current efforts by practitioners~\cite{datatoviz,datavisualizationcatalogue,datavizproject} provide hints on how this idea can be realized. For example, a decision tree can be used to find a chart based on data types or task goals. However, as mentioned earlier, practitioners' taxonomy is currently not consistent and streamlined compared to the research side. To alleviate this issue, we could utilize \textbf{templates to create more structured and easily consumable guidelines}, as demonstrated in Figure \ref{fig:template}. These templates can also be transformed into computational constraints to enhance a visualization recommendation system~\cite{moritz2018formalizing,zeng2023too}.

While building such a design guideline knowledge base is feasible given the relatively young field of visualization  the substantial ongoing challenge lies in maintaining its relevance and functionality over time. We believe that it will \textbf{require collective efforts from both practitioners and researchers}. Even if there have been efforts for both stakeholders to meet and collaborate, we have not yet seen a cohesive community that brings both together. For example, IEEE VIS primarily caters to researchers, while the Data Visualization Society focuses on practitioners. Given the challenges of creating a new community site~\cite{kraut2012building}, a viable alternative is to utilize existing communities to foster collaborations.

Practitioners often write guidelines to promote their commercial tools (e.g., DataWrapper~\cite{datawrapper}) or attract clients (e.g., Data Viz Project~\cite{datavizproject}), similar to how researchers write blog articles for recognition. Yet, interviews highlighted a lack of incentives as a major issue. Addressing this, we should \textbf{consider providing both intrinsic and extrinsic motivation} to support the knowledge base's maintenance and development. Ko's market-based incentive model~\cite{reviewmodel}, offering tokens for contributions, presents an intriguing solution. Additionally, harnessing recent advancements in large language models (LLM) could offer intelligent assistance in converting empirical results into practical guidelines~\cite{kim2023good}. For instance, LLMs can efficiently extract summary insights from research papers, translate these insights into practitioner-friendly language, and create relevant examples through code generation and contextual understanding.

The field of data visualization is not alone in facing a knowledge gap between research and practice; this is a common challenge across many disciplines. In medicine, evidence-based practices integrate clinical expertise with the best available research evidence~\cite{sackett1996evidence}. In policy-making, there is an emphasis on translating research findings into actionable policies and practices~\cite{bogenschneider2011evidence}. In education, it is common to use educational research to improve teaching and learning practices~\cite{davies1999evidence}. Learning from these established fields can also inform how the visualization field should progress to promote \textbf{evidence-based design practice}, such as developing knowledge brokering roles~\cite{ward2009knowledge} and investing collaborative research-practice partnerships~\cite{coburn2016research}. The recent VisGuides workshop also provides several insights into bridging the knowledge gap, highlighting the risks of guidelines suppressing creative solutions~\cite{visguides2022} and the development of practice-oriented guidelines~\cite{parsons2022considering}.

\section{Conclusion}
Our study investigated the guideline knowledge gap between visualization research and practice. We found a concentration of guidelines on basic chart types and task types, a lack of transfer of empirical research to practice, and a cohesive opinion on the knowledge gap among practitioners and researchers. We distilled the lessons learned by discussing various strategies to develop a sustainable visualization design guideline knowledge base and promote evidence-based design practices.

\section*{Acknowledgments}
We would like to acknowledge the support of the National Science Foundation (\#2146868). We also thank the students who assisted with the guideline-empirical study mappings, the reviewers for their valuable feedback, and the practitioners and researchers who participated in the surveys and interviews.

\bibliography{reference}
\bibliographystyle{IEEEtran}


 




\vfill

\end{document}